\pdfoutput=1
\documentclass[twocolumn]{aastex6} %% preprint2 produces a double-column, single-spaced document:
\usepackage{natbib}
\usepackage{amsmath}
\usepackage{amssymb}
\usepackage{subfigure}
\usepackage{url}
\usepackage{CJK}
\renewcommand{\vec}[1]{ {\mathbf #1} }

\newcommand{\curl}{ {\bf \nabla} \times}

\newcommand{\jcphy}{  {\it J.~Comput.~Phys.}}

\newcommand{\Fig}{{Figure}}

\shorttitle{Magnetic topology of X9.3 flare}
\shortauthors{Jiang et al.}

\begin{document}
\begin{CJK*}{UTF8}{gbsn}

  \title{Magnetohydrodynamic Simulation of the X9.3 Flare on 2017
    September 6: \\
    Evolving Magnetic Topology}
%\title{Three-Dimensional Magnetic Configuration and Evolution of a Super Solar Flare}

\author{
  Chaowei Jiang\altaffilmark{1,2,3},
  Peng Zou\altaffilmark{1},
  Xueshang Feng\altaffilmark{2,1},
  Qiang Hu\altaffilmark{3},
  Rui Liu\altaffilmark{4},
  P. Vemareddy\altaffilmark{5},
  Aiying Duan\altaffilmark{6},
  Pingbing Zuo\altaffilmark{1,2},
  Yi Wang\altaffilmark{1,2},
  Fengsi Wei\altaffilmark{1,2}}

\altaffiltext{1}{Institute of Space Science and Applied Technology,
  Harbin Institute of Technology, Shenzhen 518055, China,
  chaowei@hit.edu.cn}

\altaffiltext{2}{SIGMA Weather Group, State Key Laboratory for Space
  Weather, National Space Science Center, Chinese Academy of Sciences,
  Beijing 100190, China}

\altaffiltext{3}{Center for Space Plasma and Aeronomic Research, The
  University of Alabama in Huntsville, Huntsville, AL 35899, USA}

\altaffiltext{4}{CAS Key Laboratory of Geospace Environment,
  Department of Geophysics and Planetary Sciences, University of
  Science and Technology of China, Hefei 230026, China}

\altaffiltext{5}{Indian Institute of Astrophysics, II-Block, Koramangala, Bengaluru-560034, India}

\altaffiltext{6}{Key Laboratory of Computational Geodynamics,
  University of Chinese Academy of Sciences, Beijing 100049, China}

\begin{abstract}
Three-dimensional magnetic topology is crucial to understanding the explosive release of magnetic energy in the corona during solar flares. Much attention has been given to the pre-flare magnetic topology to identify candidate sites of magnetic reconnection, yet it is unclear how the magnetic reconnection and its attendant topological changes shape the eruptive structure and how the topology evolves during the eruption. Here we employed a realistic, data-constrained magnetohydrodynamic simulation to study the evolving magnetic topology for an X9.3 eruptive flare that occurred on 2017 September 6. The simulation successfully reproduces the eruptive features and processes in unprecedented detail. The numerical results reveal that the pre-flare corona contains multiple twisted flux systems with different connections, and during the eruption, these twisted fluxes form a coherent flux rope through tether-cutting-like magnetic reconnection below the rope. Topological analysis shows that the rising flux rope is wrapped by a quasi-separatrix layer, which intersects itself below the rope, forming a topological structure known as hyperbolic flux tube, where a current sheet develops, triggering the reconnection. By mapping footpoints of the newly-reconnected field lines, we are able to reproduce both the spatial location and, for the first time, the temporal separation of the observed flare ribbons, as well as the dynamic boundary of the flux rope's feet. 
Futhermore, 
%{Despite that the modeled reconnection is a few times faster than the observed one, 
the temporal profile of the total reconnection flux is comparable to the soft X-ray light curve. Such a sophisticated characterization of the evolving magnetic topology provides important insight into the eventual understanding and forecast of solar eruptions.
%The simulated reconnection rate is compared with the flare-ribbon separation speed as well as soft X-ray light curve.
\end{abstract}


\section{Introduction}
\label{sec:intro}

The Sun often produces major eruptive phenomena which impulsively release vast energy on the order of $10^{32}$~erg in a few minutes and strongly influence space weather. Such phenomena, observed as solar flares, filament eruptions, and coronal mass ejections (CMEs), are recognized to have a common driver, the magnetic field. This is because in the solar corona, magnetic field plays a dominant role in the plasma dynamics and solar eruptions are manifestation of sudden releases of free magnetic
energy~\citep{Aschwanden2004Book}. Magnetic reconnection, which is associated with variation of magnetic field topology, is thought to
be {the central mechanism that leads to rapid dynamical evolution that, ultimately, converts free magnetic energy into radiation, energetic particle acceleration}, and kinetic energy of
plasma~\citep{Priest2002}. Thus unraveling the magnetic configuration
of solar eruptions, in particular, the magnetic topology responsible for magnetic reconnection as well as its evolution during
flares, is essential for understanding the nature of solar eruptions.

%--review 2D model which is too simple
Without a direct measurement of the coronal magnetic field, many
theoretical models for solar eruption have been
proposed~\citep{Shibata2011} to fit observations. For instance, the
so-called standard flare model~\citep[i.e., the CSHKP flare
model,][]{Carmichael1964, Sturrock1966, Hirayama1974, Kopp1976} is
most commonly invoked. When the magnetic configuration is of concern,
this model provides simply a conceptual cartoon, in which a magnetic
flux rope (MFR) in the corona, i.e., a bundle of twisted magnetic
field lines lying above the polarity inversion line (PIL) of
photospheric magnetic field and with their legs anchored at the
photosphere, {is ejected} into the interplanetary space and forms a
CME. {Left behind, in the ejection's wake, is} an electric current sheet (CS) formed between
the stretched magnetic field lines tethering the MFR. {The ejection's rise together with magnetic reconnection in its trailing CS release the stored magnetic energy. Nonthermal particles accelerated during the release of magnetic energy} traces the newly reconnected field lines to the
chromosphere, resulting in two parallel chromospheric flare ribbons on
both sides of the PIL. {The temporal evolution of such process is
observed as a progressive separation of these two ribbons from each
other, as more and more flux reconnects}.

The full 3D magnetic configuration and dynamic evolution of solar
eruptions are mostly investigated by numerical simulation based on the
magnetohydrodynamics (MHD) model, which can describe well the
macroscopic physical behavior of the solar corona. MHD simulation of
solar eruption with idealized magnetic configuration (that is, not
directly constrained by observed magnetograms) has been performed in a
number of papers \citep[e.g.,][]{Mikic1994, Linker2003, Amari2003,
  Roussev2003, MacNeice2004, Torok2005, Fan2007, Kliem2010,
  Aulanier2010, Kusano2012, Torok2013, Wyper2017, MeiZ2018}. In
particular, the 3D magnetic topology evolution of an MFR eruption has
been extensively investigated in a series of
papers~\citep{Aulanier2010, Aulanier2012, Janvier2014, Janvier2015}.
It is found that the MFR is wrapped around by a topological
quasi-separatrix layer (QSL) which separates the rope {from} its ambient
flux. Such QSL consists of a continuous set of sheared magnetic field
lines that defines the boundary surface of the rope, and reconnection
occurs mainly below the rope, between the sheared arcades in a
{tether-cutting form~\citep{Moore2001}}, or strictly speaking, slipping
reconnection, see \Fig~4 of~\citep{Aulanier2010}. The reconnection site is
actually an intersection of the QSL with itself below the MFR, which
{forms} a hyperbolic flux tube \citep[HFT,][]{Titov2002} and its 2D cross
section corresponding to a X point configuration.  The photospheric
footprints of the QSL display two J-shaped ribbons, with the legs of
the MFR anchored within the hooked parts. Thus reconnection in the QSL
produce flare ribbons of J shapes, and with the separation of the main
part of the ribbons, the hooks also expand.  If the MFR is highly
twisted, the hooks would close onto themselves, form two rings as
predicted by theoretical model~\citep{Demoulin1996} but is not
reproduced by the numerical model of~\citet{Aulanier2010}. Observation
indeed shows such closed-ring-shaped ribbons connecting the ends of
the two main ribbons. For instance, \citet{WangW2017} observed two
closed-ring-shaped flare ribbons in the case of buildup of
high-twisted MFR with the development of a flare reconnection. During
the separation of the main flare ribbons, the flare rings expand
significantly starting from almost point-like
brightening. Furthermore, transient coronal holes, i.e., post-eruptive
coronal dimmings, are naturally suggested to map the feet of eruptive
MFRs, along which mass {leakage} into interplanetary space could take
place~\citep{Qiu2007, Webb2000}.

Idealized MHD simulations are also commonly used to investigate the
initiation mechanism of eruptions. Two kinds of ideal MHD
instabilities have been commonly invoked as being the main driver of
the eruption of an MFR. The first one, kink
instability~\citep[KI,][]{Hood1981} depends on the twist degree of the
magnetic field line in the rope. MHD simulations suggest that KI occur
when the number of turns in the field lines around the rope axis
exceeds critical value of $1.75 \sim 2$~\citep{Torok2004,
  Fan2003}. Such a highly wound flux rope then evolves to reduce this
strong internal twist by transferring some of its twist into writhe
(deformation of the rope axis), conserving helicity in the
process. The second one is the torus instability~\citep[TI,
][]{Kliem2006PRL, Myers2015}, which is a result of the loss of balance between
the ``hoop force'' of the rope itself and the ``strapping force'' of
the ambient field. The TI is determined by a decay index of the
strapping field, which quantifies the decreasing rate of the
strapping force along the distance from the torus center. It is found
that the TI occurs when the apex of the rope enters into a domain with
decay index larger than a threshold of $\sim 1.5$ based on theoretical
studies~\citep{Kliem2006PRL} as well as MHD
simulations~\citep{Fan2007, Aulanier2010}.

%There are competing models which stress the primary role of magnetic reconnection, while the MFR is not a necessity for triggering and driving eruptions. It is proposed in a tether-cutting flare model~\citep{Moore2001} that flare can be triggered by reconnection between strongly sheared magnetic arcades near their inner footpoints. Such reconnection will results in newly-formed magnetic field line expanding upward by magnetic tension force, and an eruption is presumably triggered by a positive feedback between the reconnection and the upward expansion. An alternate to the reconnection below the sheared arcades is reconnection above them, which is proposed in the breakout eruption model~\citep{Antiochos1999}. The breakout model requires a quadrupolar magnetic configuration to form a coronal null point or similar structure that is favorable for reconnection to remove the overlying flux that straps the sheared field, which can then break out to form eruption. In these reconnection-dominated models, twisted MFR does not exist prior to the eruption, while it can be built up during the eruption.

%~讲数据驱动、数据约束方法来刻画更实际的太阳爆发过程 举例子。
Realistic simulation of solar eruptions constrained or driven by
photospheric magnetograms (and other observable features) provides a
significant step forward in understanding the complexity of magnetic
configuration and evolution in real events. The power of such
kind of simulations has been demonstrated by many authors~\citep[see a
review by][]{Inoue2016Review}. For instance, \citet{Jiang2013MHD}
simulated the sigmoid eruption in AR~11283, which possesses a complex
configuration consisting of a MFR and a spine-fan null-point topology
linking to multiple polarities. They first reconstructed an
approximately NLFFF model for the instant immediately prior to the
eruption and found that the magnetic field is
unstable~\citep{Jiang2013MHD, Jiang2014NLFFF}. Then the unstable field
is used to initialize a full MHD simulation, which can reproduce the
subsequent eruption in remarkably agreement with the observed filament
ejection. \citet{Jiang2016NC} further developed a data-driven MHD model
which self-consistently follows the time-line of a flux-emerging AR
over two days leading finally to a eruption. \citet{Kliem2013} studied
the eruption on 2010 April 8 by initializing their zero-$\beta$ MHD
model with an unstable pre-flare field model constructed by a
flux-rope insertion technique~\citep{Ballegooijen2004,
  Ballegooijen2007}. It also yields good agreement with some observed
features.  \citet{Inoue2014} investigated the eruption mechanism of an
X2.2 flare in the well-known AR~11158~\citep{Sun2012}. They first
extrapolated an NLFFF model using SDO/HMI vector magnetogram observed
two hours before the flare, and found this NLFFF is stable in MHD
simulation. Thus, an enhanced anomalous resistivity was used to
increase the magnetic twist through tether-cutting reconnection in
those sheared arcades in the AR core, after which the
quasi-equilibrium was broken and an eruption followed.  Similar
approaches are also adopted in~\citep{Amari2014nat, Amari2018,
  Inoue2018}.

Based on these data-constrained and data-driven simulations, the 3D
magnetic topological evolution and its relation with observed flare
ribbons were investigated recently.  Using the flux-rope insertion
method, \citet{Savcheva2015} modeled the magnetic field of seven
two-ribbon flares and found that the main ribbons are matched well by
the flux-rope-related QSLs except some parts of the hooks of the
J-shaped ribbons. \citet{Savcheva2016} further studied the evolution
of unstable flux-rope models using their magneto-frictional code and
showed that the evolution of flare ribbons can also partially
reproduced by the tracking the evolution of the flux-rope-related
QSLs.  In the data-driven simulation of a flux-emergence process which
leads finally to eruption, \citet{Jiang2016NC} found that during the
emergence, a null-point-like magnetic topology is formed with a
quasi-circular QSL whose footprints match the observed quasi-circular
flare ribbon. The same data-driven model was used to study the great
confined flare of X3.1 in super AR~12192~\citep{Jiang2016ApJ}, and strikingly good match of
the reconnecting field-line footpoint with flare ribbons was achieved, but only a
time snapshot was shown. Very recently, \citet{Jiang2017} modeled the
magnetic field of a peculiar X-shaped-ribbon flare and found that there a large-scale
current sheet existing prior to the flare and the footpoints of field lines tracing from the CS reproduce
the shape of the ribbons.

This paper is devoted to a comprehensive analysis of the 3D magnetic configuration and evolution
of a great eruptive flare occurred on 2017 September 6 with a data-constrained MHD simulation.
The studied flare, reaching
GOES X9.3 class, was the largest one of the last decade, and quickly
drew intensive attentions in the communities of
solar physics~\citep{YangS2017, SunX2017, Warren2017,
  YanX2018, LiY2018, WangH2018, HuangN2018} as well as space weather~\citep[e.g.,][]{LeiJ2018}. The flare occurred in a magnetic complex due to the interaction of
multiple magnetic polarities as observed on the photosphere.
Our MHD simulation realistically reproduces the
dynamic evolution of the magnetic field underlying the flare.
%, which is strongly impacted by such eruptive flares.
In particular, we focus on the magnetic topology and its evolution during the flare.
{Our model of the field at the time of the X9.3 flare contains a complex and unstable MFR system}, which is possibly due to TI, and the eruption is
resulted by the consequent expansion of the MFR. Furthermore with an
accurate topology analysis, we find that the footprint of those field
lines reconnected underlying the MFR {roughly} matches both the spatial location
and its temporal evolution of flare ribbons. We will first describe data and models in Section~\ref{sec:model}, then present the simulation results as well as its comparison with
observations in Section~\ref{sec:result}, and finally conclude in
Section~\ref{sec:concl}.

% all the figures------------------------------------------------------------------------

\section{Data and Models}
\label{sec:model}

\subsection{Event and Data}

The investigated flare \texttt{SOL2017-09-06T11:53}, {which is the largest flare in solar cycle~24}, took place in a super flare-productive solar AR, NOAA 12673. In this AR,
4 X-class and 27 M-class flares are produced from 2017 September 4 to
10. The X9.3 flare on September 6 started at 11:53~UT, impulsively
reached its peak at 12:02~UT and then ended at 12:10~UT, and also accompanied by a large CME~\citep{YanX2018}. Its location
on the solar disk is shown in \Fig~\ref{fig1}, as imaged by the
Atmospheric Imaging Assembly (AIA) onboard the Solar Dynamics
Observatory (SDO). The SDO/AIA can provide a full-disk image of the Sun simultaneously in 6 EUV filters, including 171~\AA, 193~\AA, 211~\AA, 335~\AA, 94~{\AA}, and 131~\AA. The spatial resolutions of all these filters are $0.6$~arcsec and the cadences are 12 seconds.

When AR~12673 rotated to the solar limb on 2017
September~10 (\Fig~\ref{fig1}c), it produced an X8.2 flare, which is
the second largest one after the X9.3 flare. As the two flares are
generated in the same region, {they might plausibly have basically the same} 3D magnetic configuration. Thus the observation of this limb
flare provides a side view of the 3D structure underlying the flares,
in addition to the nearly top view for the X9.3 flare.
The AIA image of this limb flare will be used to compare
qualitatively with our simulation of the X9.3 flare in a 2D slice.

The vector magnetogram used for our coronal field extrapolation is
taken by the Helioseismic and Magnetic Imager~\citep[HMI,
][]{Schou2012HMI} on board {\it SDO}. In particular, we used the data
product of the Space-weather HMI Active Region Patch~\citep[SHARP,
][]{Bobra2014}, which has been resolved 180$^{\circ}$ ambiguity by
using the minimum energy method, modified the coordinate system via
the Lambert method and corrected the projection effect. {The magnetogram for this AR is well flux-balanced as the ratio of the total flux to the total unsigned flux is $\sim 0.05$.} Since this flare was associated with an erupting filament, the
H$\alpha$ data with spatial resolution of 1~arcsec from Global Oscillation Network Group (GONG) are used as well for checking the location of the filament.

\subsection{NLFFF Model}

The pre-flare coronal magnetic field is extrapolated by our
CESE--MHD--NLFFF code~\citep{Jiang2013NLFFF}. It belongs to the class
of MHD relaxation methods that seek approximately force-free equilibrium
\begin{equation}\label{nlfff}
  (\nabla \times \vec B)\times \vec B = \vec 0
\end{equation}
for given boundary value specified by observed vector magnetograms.
It solves a set of modified zero-$\beta$ MHD equations with a friction force using an
advanced conservation-element/solution-element (CESE) space-time
scheme on a non-uniform grid with parallel
computing~\citep{Jiang2010}. Starting from a potential field
extrapolated from the vertical component of the vector magnetogram,
the zero-$\beta$ MHD system is driven to evolve by incrementally changing the
transverse field at the bottom boundary until matching the vector
magnetogram, after which the system will be relaxed to a new
equilibrium.
{In the code, a pseudo plasma density $\rho=B^2$ is used in the momentum equation. We use two terms in the induction equation to control the nonzero magnetic divergence, one is the Powell source term $-\vec v \nabla \cdot \vec B$~\citep{Powell1999}, and the other is a diffusion term $\nabla(\mu \nabla \cdot\vec B)$ where $\mu$ is the diffsion coefficient. These two terms can effectively control the numerical magnetic divergence.}
The code has an option of using adaptive mesh refinement
and multi-grid algorithm for optimizing the relaxation process~\citep{Jiang2012apj}.
The computational accuracy is further improved by a magnetic-field
splitting method, in which the magnetic field is divided into a
potential-field part and a non-potential-field part and only the
latter is actually {evolved in the MHD relaxation to derive the NLFFF field.} Before being input into the
code, the raw vector magnetogram is required to be preprocessed
to reduce the Lorentz force it contain. Furthermore, to be
consistent with the code, we developed a unique preprocessing
method~\citep{Jiang2014Prep} that also splits the vector magnetogram
into a potential part and a non-potential part and handles them
separately. Then the non-potential part is modified and smoothed by an
optimization method similar to ~\citet{Wiegelmann2006} to fulfill the
conditions of total magnetic force-freeness and
torque-freeness. {
The preprocessing alters the original HMI magnetogram in all three components. A simple way to convey the extent of the changes is to compute planar (2D) versions of the quantities defined by Equations (28)-(31) in~\citet{Schrijver2006}. These are: $C_{\rm vec}=0.95$; $C_{\rm cs}=0.69$; $E_{\rm n}=0.38$; $E_{\rm m} =0.72$. Ideally, the first two would be $1.0$, and the latter would be zero. These discrepancies are mostly due to the smoothing of the data.
Details of the CESE--MHD--NLFFF code and the preprocessing method are described in a series of
papers~\citep{Jiang2013MHD, Jiang2014Prep, Jiang2014formation, Jiang2014NLFFF}}. It is
well tested by different benchmarks including a series of
analytic force-free solutions~\citep{Low1990} and numerical
MFR models~\citep{Titov1999, Ballegooijen2004}, and have been applied to the SDO/HMI vector
magnetograms~\citep{Jiang2013NLFFF, Jiang2014NLFFF}, which enable to reproduce
magnetic configurations in very good agreement with corresponding
observable features, including coronal loops, filaments, and sigmoids.

\subsection{MHD Model}
\label{sec:mhd}
The MHD simulation is realized by solving the full set of 3D,
time-dependent ideal MHD equations with solar gravity. The initial
condition consists of the magnetic field provided by the NLFFF model and a
hydrostatic plasma. {While the NLFFF derivation procedure only solved for $\vec B$ and $\vec v$ in a pseudo-evolution, in the MHD model of the eruption, all MHD variables $(\rho, \vec v, \vec B, p)$ are solved for.}
The initial temperature is uniform, with a value
typically in the corona, $T=10^{6}$~K (which gives sound speed
$c_{S}=128$~km~s$^{-1}$). The initial plasma density is uniform in
horizontal direction and vertically stratified by the gravity. To
mimic the coronal low-$\beta$ and highly tenuous conditions, the
plasma density is configured to make the plasma $\beta$ less than
$0.1$ in most of the computational volume. The smallest value of
$\beta$ is $5\times 10^{-4}$, corresponding to the largest Alfv{\'e}n
speed $v_{\rm A}$ of approximately $8$~Mm~s$^{-1}$. The units of length and time in
the model are $L=11.5$~Mm (approximately 16~arcsec on the Sun disk) and
$\tau=L/c_{S}=90$~s, respectively. {In this simulation, the MHD code was run for 1~$\tau$.} The MHD solver is the same CESE
code described in \citet{Jiang2010}. We use a non-uniform grid
with adaptive resolution based on the spatial distributions of the magnetic field and
current density in the NLFFF model. This grid is designed for the sake
of saving computational resources without losing numerical accuracy,
and more details of this can be found in \citep{Jiang2017}. The
smallest grid is $\Delta x = \Delta y = 2\Delta z=0.36$~Mm
(approximately 0.5~arcsec on the Sun).  A moderate viscosity $\nu$, which
corresponds to Reynolds number $R_e = L v_{\rm A} /\nu$ of $\sim 10^2$, is used to keep the
numerical stability of the code running for the whole duration of the flare eruption process. No explicit resistivity is
included in the magnetic induction equation, and magnetic reconnection
is still allowed due to numerical resistivity $\eta$, which corresponds to the Lundquist number (or magnetic Reynolds number) of $S = L v_{\rm A}/\eta $ of $\sim 5\times 10^3$ in our grid settings and numerical scheme.
%if any CS forms and becomes thin enough with thickness close to the grid resolution (i.e., the smallest grid).
Although there is no doubt that the viscosity and
numerical resistivity in our model overestimate the real values in the
coronal plasma (which are on the order of $10^8\sim 10^{10}$), the basic magnetic topological evolution as simulated
is still robust~\citep[see also][]{Jiang2016NC}.
The computational volume is slightly larger than the size of simulated AR, and the simulation is stopped before any disturbance
reaches the numerical boundaries. At the bottom boundary (i.e., the coronal base),
all the variables are fixed ({thus, the density and temperature are constant in both time and space, and the
velocities are held at zero}) except the transverse components of the
magnetic field, which are released (or floated) by linear extrapolation from the inner points along the $z$-axis.

In the combination of NLFFF model and MHD simulation, it should be
noted that almost all the available NLFFF codes actually generate
non-force-free magnetic field data with residual Lorentz forces that
 are often non-negligible. {In our code, some of the residual forces in this NLFFF procedure will arise from the artificial friction used in the method.} The magnitude of these residual forces can indicated by the misalignment of the
current $\vec J$ and magnetic field vector $\vec B$~\citep{Schrijver2006},
which is usually measured by CWsin, a current-weighted average sine of the
angle between $\vec J$ and $\vec B$. CWsin is typically in the range
of $0.2$ to $0.4$~\citep[see, e.g.,][]{Schrijver2008a, DeRosa2009,
  DeRosa2015}. Another metric $E_{\curl \vec B}$ measuring more
directly the residual Lorentz force is defined as the average ratio of
the force to sum of the magnitudes of magnetic tension and pressure
forces~\citep{DuanY2017}. In the studied event here, these two metrics
for the CESE--MHD--NLFFF extrapolation are respectively, CWsin $=0.23$
and $E_{\curl \vec B} = 0.17$. These
metrics are reasonably small as compared with other codes~\citep[e.g.,
see the last column of Table~2 in][]{DeRosa2015}, but such residual force can instantly induce plasma motion in a low-$\beta$ and highly tenuous plasma environment.
Actually, this initial motion provides a way of perturbing
the system. If the system is very stable, i.e., significantly far away from an unstable regime, it will quickly relax to MHD equilibrium as the
induced motion can alter the magnetic field, which in turn generate
restoring force to brake the motion. Otherwise, if the system is
unstable or not far away from an unstable regime, the perturbation could grow
and lead to a drastic evolution of the system as driven by the instability. Thus the combination of NLFFF and MHD model can be used to test the potentially unstable nature or instability of numerical NLFFF, {while here we cannot assess what is exactly the mechanisms that make the extrapolated field unstable}.
%Although such combination of NLFFF and MHD models might not be able to identify the true mechanism triggering flare,
Nevertheless, it still provides a viable tool to reproduce the fast magnetic evolution during the flare.

\subsection{Magnetic Field Analysis Method}

We used a set of magnetic field analysis methods including search of magnetic bald patches~\citep[BPs,][]{Titov1993}, calculation of magnetic twist number, squashing degree and decay index, which are described in the following.

BPs are places on photospheric PIL where the transverse field directs from the negative polarity to positive one. This is inverse to a normal case that transverse field directs from positive flux to negative one, and thus the field line is concave upward. BPs are special because they defines a magnetic topology separatrix, known as BP separatrix surface~\citep[BPSS,][]{Titov1999}, which are often associated with MFR that is attached with the photosphere. BPs can be located by searching the point on the magnetogram where the conditions
\begin{equation}\label{eq1}
  \vec B\cdot \nabla B_z > 0,\ \ B_z=0
\end{equation}
are satisfied. Magnetic dips are searched using the same conditions but applied for the full 3D volume of the field.

The magnetic twist number $T_w$ for a given (closed) field line is defined by~\citep{LiuR2016}
\begin{equation}\label{Tw}
  T_w=\int_L \frac{(\nabla \times \vec B)\cdot \vec B}{4\pi B^2} dl
\end{equation}
where the integral is taken along the length $L$ of the magnetic field line from one footpoint to the other. Precisely, $T_w$ measures the number of turns two infinitesimally close field lines wind about each
other~\citep{LiuR2016}.

The squashing degree $Q$ is derived based on the mapping of two footpoints for a field line. Specifically, a field line starts at one footpoint $(x,y)$ and ends at the other footpoint $( X(x,y), Y(x,y) )$. Then the squashing degree associate with this field line is given by~\citep{Titov2002}
\begin{equation}
  \label{eq:Q}
  Q = \frac{a^{2}+b^{2}+c^{2}+d^{2}}{|ad-bc|}
\end{equation}
where
\begin{equation}
  a = \frac{\partial X}{\partial x},\ \
  b = \frac{\partial X}{\partial y},\ \
  c = \frac{\partial Y}{\partial x},\ \
  d = \frac{\partial Y}{\partial y}.
\end{equation}
Usually QSLs can be defined as locations where $Q>>2$.

In the torus instability (TI)~\cite{Kliem2006PRL}, which is a result of the loss of balance between the ``hoop force'' of the rope itself and the
``strapping force'' of the ambient field, the decay index $n$ plays a key role. It quantifies the decreasing
{strength} of the strapping force along the distance from the torus
center.
Here $n$ is calculated in the vertical cross section perpendicularly crossing the main axis of the rope~(see \Fig~\ref{fig3}c) in such manner: we regard the bottom PIL point, named O (denoted by the black circle in the figure) as the center of the torus, and for a given grid point P, $n({\rm P})=-d \log (B_{\rm p})/ d \log(h)$,  where $B_{\rm p}$ is the magnetic field component perpendicular to the direction vector $\vec r_{\rm OP}$, and $h = |\vec r_{\rm OP}|$. Here the strapping field is approximated by the potential field model that matches the $B_z$ component of the photospheric magnetogram~\citep{Aulanier2010, Jiang2013MHD}. The TI occurs when the apex of the rope
enters into a domain with decay index larger than a threshold of $\sim
1.5$ based on theoretical studies~\citep{Kliem2006PRL}.

\section{Results}
\label{sec:result}

\subsection{Magnetic field on the photosphere}

\begin{figure*}[htbp]
  \centering
  \includegraphics[width=\textwidth]{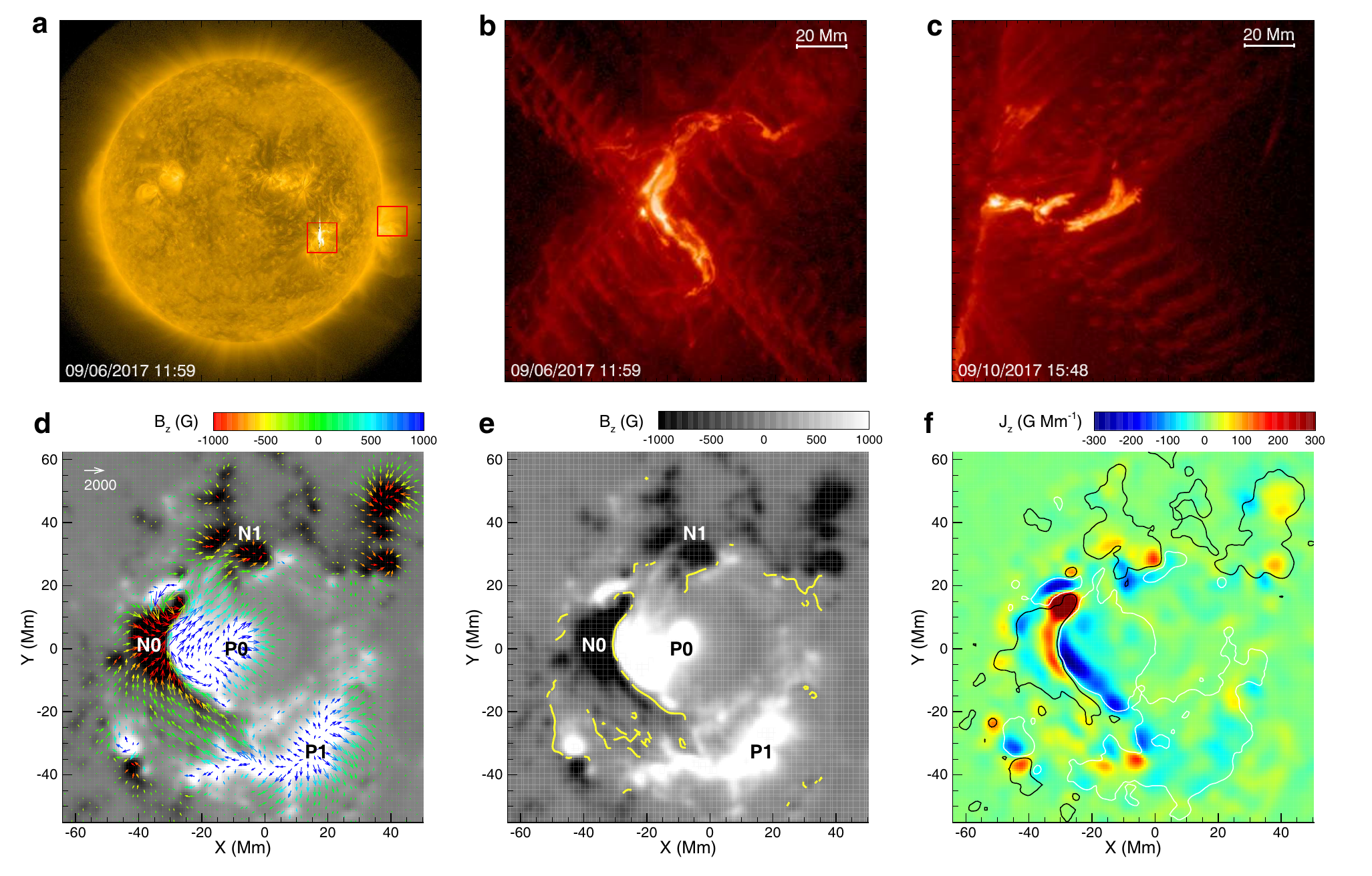}
  \caption{
  {The flare location and photospheric magnetic field.}
  (a) Full-disk image of the Sun observed in SDO/AIA 171~{\AA}. The two boxes indicate the locations of the on-disk X9.3 flare on September 6 and the limb X8.2 flare occurred on September 10. (b)-(c) SDO/AIA 304~{\AA} image of the X9.3 flare and the X8.2 flare, respectively.
  (d)-(f) SDO/HMI vector magnetogram taken at 11:36~UT on September 6, which
  is 17~min before the X9.3 flare onset. In (d), the magnetic flux distribution, i.e., $B_z$, is overlaid by the transverse field vector $(B_x, B_y)$ as denoted by the colored arrows. The main magnetic polarities P0, N0, P1, and N1 are
  labeled. In (e) the yellow curves are the BP locations along the PIL. (f)
  Distribution of the vertical current density, which is defined as $J_z = \partial_x B_y - \partial_y B_x$. The contour lines are plotted for $B_z = -500$~G (colored as black) and 500~G (white). The ratio of the direct current (DC) to the return current (RC) for the positive flux is $|{\rm DC}/{\rm RC}|^+ = 2.31$, and for the negative flux is $|{\rm DC}/{\rm RC}|^- =2.26$.}
\label{fig1}
\end{figure*}

First, we analyzed the pre-flare magnetic field in the photosphere taken by SDO/HMI at the time of 11:36~UT, just 17~min ahead of the flare onset. This vector magnetogram provides the only input to our numerical models. As shown in \Fig~\ref{fig1}d, there are mainly 4 magnetic concentrations.
In the core region, two closely touched magnetic concentrations of opposite polarities, P0 and N0, are separated by a polarity inversion line (PIL) of C shape (referred to as the main PIL hereafter), nearly enclosed by another two concentrations (P1 and N1) in the south and north, respectively.
Analysis of the time-sequence magnetograms suggested that such configuration is
formed by several groups of extremely fast emerging flux blocked by a pre-existing sunspot~\citep{YangS2017}, which results in a strongly distorted magnetic system. Significant magnetic shear can be seen along the main PIL, which is so strong that magnetic BPs form on almost the whole PIL (\Fig~\ref{fig1}e). The presence of
BPs means that magnetic field lines immediately above the PIL do not connect P0-N0
directly but are concave upward, grazing over the PIL and forming magnetic dips. Such magnetic-sheared configuration with BPs is often found in the case of theoretical models of
coronal MFR that is partially attached at the photosphere~\citep{GibsonFan2006, Aulanier2010}.

Strong current can be seen directly from the transverse magnetic field. For example,
the transverse magnetic vectors form a distinct vortex
in the north end of N0, indicating strong current and magnetic twist there.
Indeed, distribution of enhanced electric currents with inverse
directions on two sides of the main PIL is derived through Amp{\`e}re's law from the transverse
magnetic field (\Fig~\ref{fig1}f), which indicates
that volumetric current channels through the
corona like a closed circuit~\citep{Janvier2014, SunX2015}.
%and such pattern of current distribution is also found in other eruptive %regions that are suggested to be involved MFR eruption.
The current is significantly non-neutralized with respect to magnetic flux of either sign:
the ratio of the direct current (DC) to the return
current (RC) for the positive flux is $|{\rm DC}/{\rm RC}|^+ = 2.31$,
and for the negative flux is $|{\rm DC}/{\rm RC}|^- =2.26$.
Such non-neutralized current has been recently recognized to be a common feature of many
eruptive ARs~\citep{LiuY2017, Kontogiannis2017, Vemareddy2017}, and can support that MFR exists prior to eruption~\citep{Torok2014}.
All these features suggest that a twisted MFR exists in the region and can likely account for the eruptive activities.

%\begin{figure*}[htbp]
%  \centering
%  \includegraphics[width=\textwidth]{fig8}
%  \caption{
%  {Strong-to-weak shear transition in post-flare loops and enhancement of horizontal field.}
%  (a) The closed magnetic arcades immediately below the reconnection site. Note that initially ($t=0$) they are long field lines that form the pre-flare MFR. The colors of the lines denote the value of height $z$. The background is shown with $B_z$ on the bottom. (b) Distribution of the horizontal field $B_h = \sqrt{B_x^2+B_y^2}$ on the bottom. The PIL of the magnetic flux $B_z$ is shown by the white-colored lines.}
%  \label{fig8}
%\end{figure*}

\subsection{Pre-flare Coronal Magnetic Configuration}

\begin{figure*}[htbp]
  \centering
  \includegraphics[width=\textwidth]{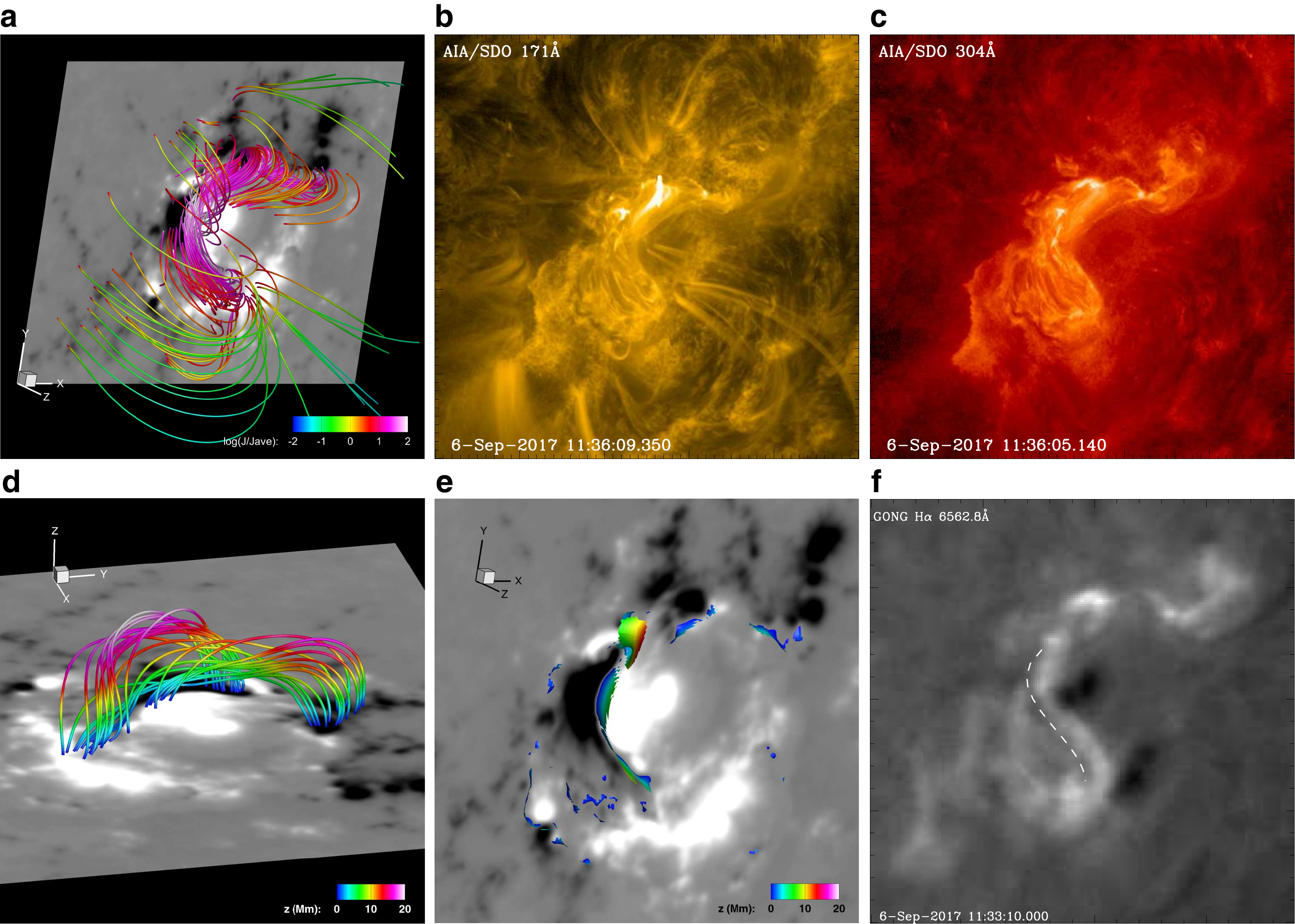}
  \caption{
  {Comparison of the reconstructed magnetic field with the observed features of the solar corona prior to the flare.}
    (a) SDO view of sampled magnetic field lines of the NLFFF reconstruction. The color of the lines represents the value of current density $J$ (normalized by its average value $J_{\rm ave}$ in the computational volume). The background is the photospheric magnetogram.
    (b) and (c) SDO/AIA 171~{\AA} and 304~{\AA} images of the pre-flare corona.
    (d) The low-lying magnetic field lines in the core region. The field lines are color-coded by the value of height $z$. (e) Locations of dips in the magnetic field lines, and the color indicates the value of height $z$.
    (f) GONG H$\alpha$ image of the AR. The dashed curve denotes the location of a long filament.}
  \label{fig2}
\end{figure*}

The coronal magnetic field in a quasi-static state prior to flare can
be well approximated by force-free models.
From the HMI vector magnetogram, our NLFFF
code reconstructs magnetic field lines that
nicely match the observed coronal loops (see \Fig~\ref{fig2}). The
pre-flare magnetic configuration comprises a set of strongly-sheared
(current-carrying) low-lying (heights of $\sim 10$~Mm) field lines in
the core region, which is enveloped by less sheared
arcades (\Fig~\ref{fig2}a and d). The low-lying field lines extend their ends to the south and north polarities P1 and N1, forming an overall C shape.
Concave-upward portions of these field lines, also termed
magnetic dips, are able to support dense filament material against the solar gravity.
As shown in \Fig~\ref{fig2}e, the dips are distributed
almost all the way along the main PIL, which is in consistence with
the distribution of BPs. They can thus support a long filament along
the PIL. Such a filament appears to exist as seen in the GONG H$\alpha$
image (\Fig~\ref{fig2}f), and the shape of the dips matches rather well with the filament.

\begin{figure*}[htbp]
  \centering
  \includegraphics[width=0.8\textwidth]{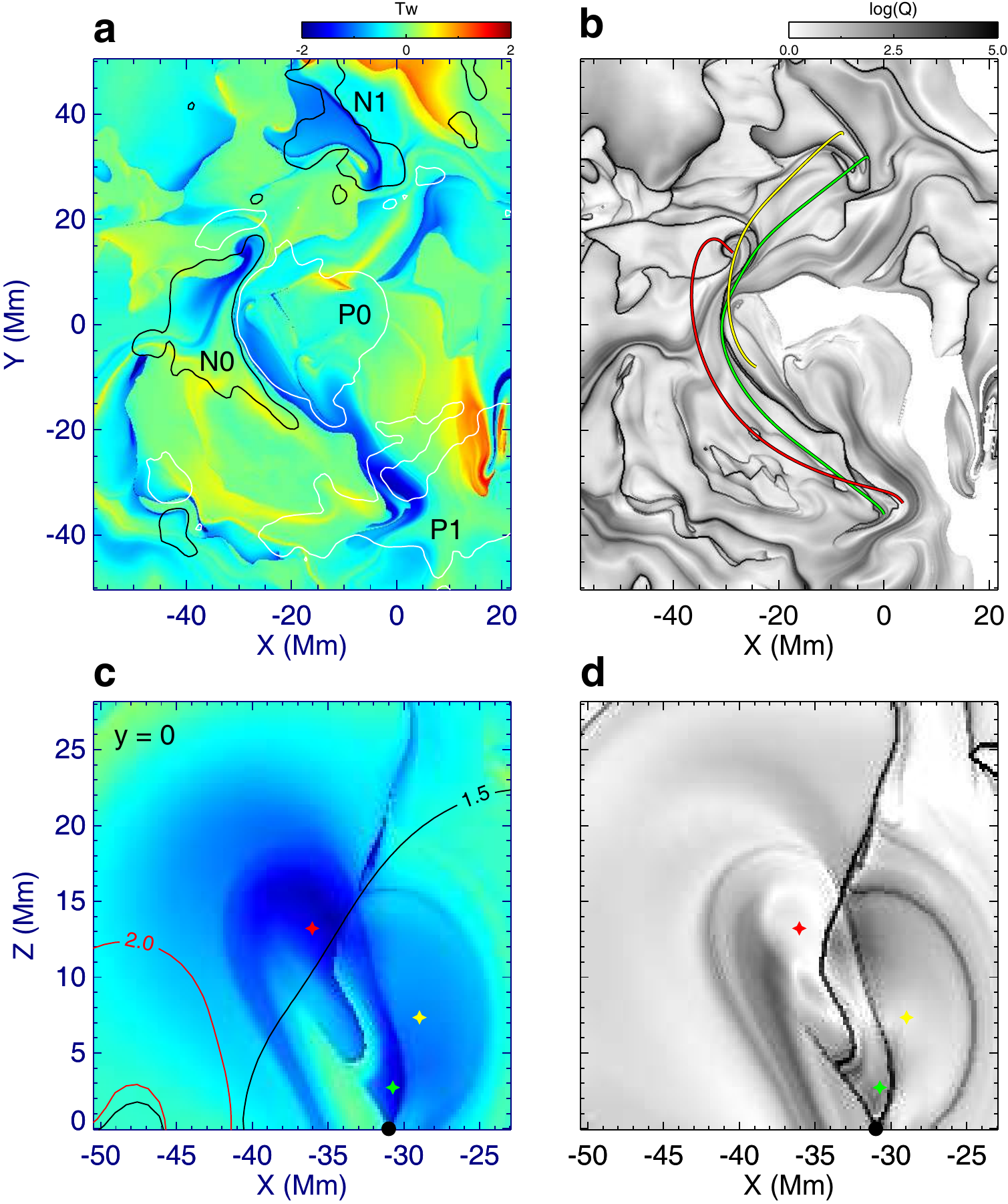}
  \caption{
  {Detailed configuration of the reconstructed pre-flare magnetic field.}
  (a) Map of magnetic twist number $T_w$ at the bottom surface $z=0$. Overlaid are contour lines for $B_z=500$~G (white) and $-500$~G (black). Coordinates are the same as shown in \Fig~\ref{fig1}d.
  (b) Map of magnetic squashing factor $Q$ at the bottom. The black thin lines as formed by the large-$Q$ value are locations of magnetic topology separatries and QSLs where the magnetic field-line mapping is
  discontinuous or {rapidly} changes. Three field lines with different colors are plotted to represent the magnetic flux of different connections, which make up the MFR.
  (c) Twist number distribution on a vertical cross section ($y=0$). The three colored stars denote the intersection points of the sample field lines shown in (b) with the cross section. The black circle indicates the main PIL. The contour lines are shown for the decay index $n=1.5$ and 2. {(d) Distribution of magnetic squashing factor $Q$ on the same cross section shown in (c).}}
  \label{fig3}
\end{figure*}

%--twist degree and qsl
The existence of MFR is confirmed by the NLFFF model. Calculation of
the magnetic twist number $T_w$ shows that the low-lying core field
is twisted left-handedly (see \Fig~\ref{fig3}). The magnetic twist
number ($T_w < 0$) is significantly enhanced along the two sides of
the main PIL, nearly in the same locations of intense
current density (compare \Fig~\ref{fig1}f and \Fig~\ref{fig3}a). The
regions of enhanced negative twist also extend to the far-side
polarities P1 and N1. Magnetic twist number in most of these regions is around 1.5,
which is close to the threshold
of KI for an idealized MFR~\citep{Torok2004}. Here the twisted
magnetic flux constitutes a complex flux-rope configuration with
{multiple domains of connectivity}.  The different connections can be
distinguished from a map of magnetic squashing degree (the $Q$
factor)~\citep{Demoulin1996, Titov2002}, which precisely maps the
topological separatries or quasi-separatrix layers (QSLs) of the
field-line connectivity. As shown in \Fig~\ref{fig3}b, this twisted
flux bundle consists of mainly three types of connections: P1-N0, P0-N1
and P1-N1, while the connection of P0-N0 does not form along the main PIL,
as a natural result of the existence of the BPs. Note that the flux connection of P1-N1
is bounded at each footpoint by a closed high-$Q$ line, the footprint of the QSL wrapping around the flux rope (\Fig~\ref{fig3}b).
In a vertical cross section (\Fig~\ref{fig3}d), this QSL touches the bottom surface, also because of the presence of BPs there.
A true separatrix, i.e., BPSS, exists at the center of this QSL where $Q \rightarrow \infty$. As shown by distribution of twist number on
the cross section $y=0$~(\Fig~\ref{fig3}c),
the MFR is non-uniformly twisted (for instance, part of the P1-N1 is has twist of only $\sim 0.5$), nor is it a coherent structure without a well-defined rope axis. We note that such multiple connections
and inhomogeneous twist are not characterized by any
theoretical (or idealized) models of MFR, but are often found in NLFFF reconstructions~\citep[e.g.,][]{Awasthi2018}.

%--decay index and TI
We further compute the decay index $n$ of the strapping field, which
is the key parameter deciding the TI of an MFR system.
In the idealized model, the TI occurs once the apex of rope axis reaches the
domain of $n>1.5$~\citep{Kliem2006PRL, Aulanier2010}, but here there is no a well-defined axis.
Even though, it is found that the major part of the MFR (i.e.,
the flux with $|T_w| > 1$) reaches a region with $n>1.5$~(see
\Fig~\ref{fig3}c). This indicates that the
MFR is already in an unstable regime, suggesting that the eruption is
more likely a result of TI rather than KI.
{However, conclusion cannot be made here because the TI (and KI) theory is derived from idealized MFR configurations, while our realistic coronal field is much more complex. Thus, we note that it is still unclear what is exactly the unstable nature of the pre-flare field.}

%
%Finally two non-potential parameters are worthy mentioned here.
%The magnetic free energy is $9.81\times 10^{32}$~erg, which is more than 60\% of
%the corresponding potential energy $1.48\times 10^{33}$~erg. This is significantly larger
%than many previous studied flare regions. On the other hand, the relative magnetic helicity, which
%quantifies magnetic twist in a whole volume, is
%$-4.71\times 10^{43}$~Mx$^{2}$. Its normalized value by the square of total unsigned magnetic flux
%is $-1.69\times 10^{-2}$, which seems to rather regular among flare regions.

\subsection{The Eruptive Evolution}

\begin{figure*}[htbp]
  \centering
  \includegraphics[width=\textwidth]{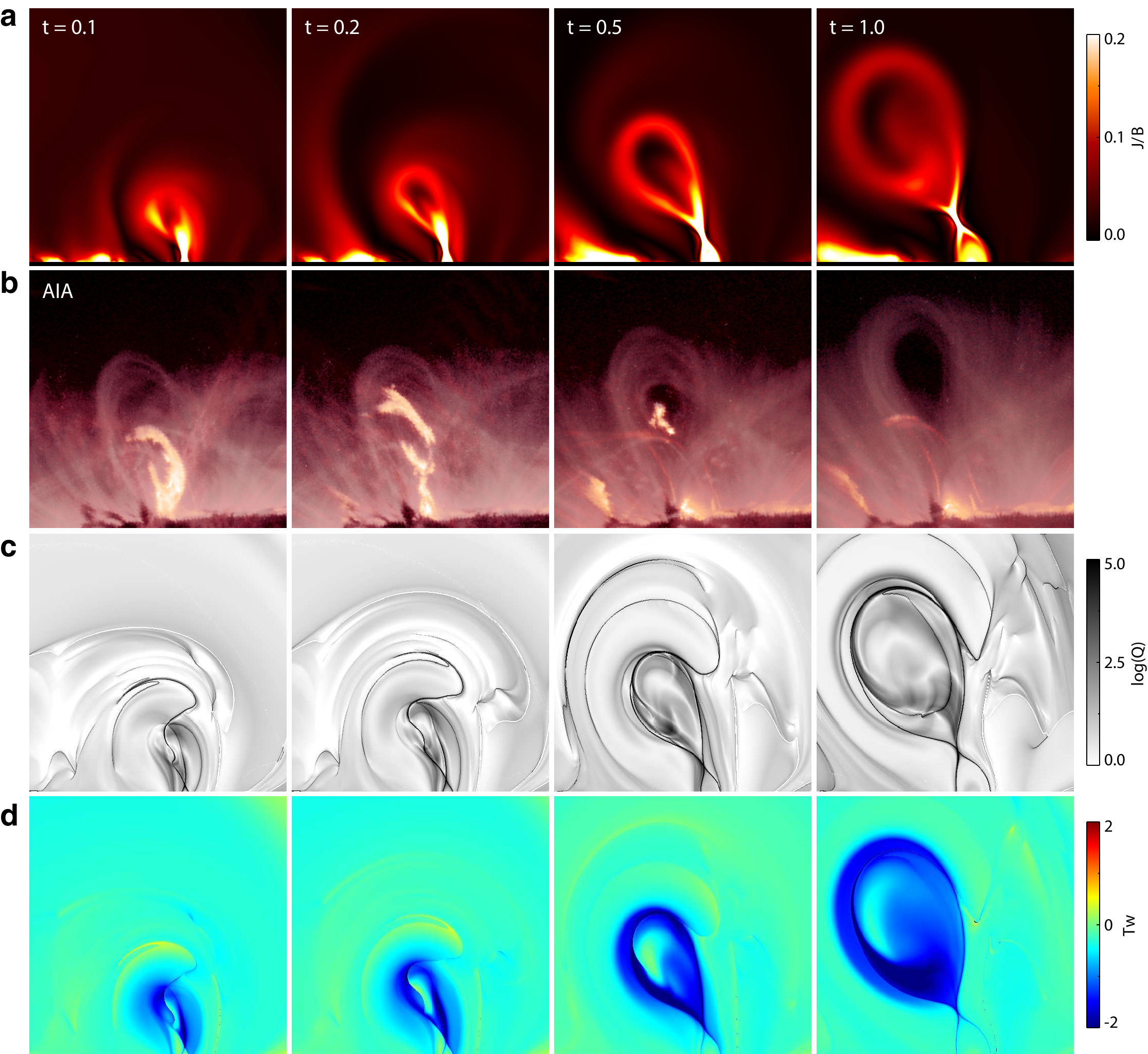}
  \caption{
  {Temporal evolution of the eruptive structure in 2D view.}
  (a) Distribution of current density on the vertical cross section (the $y=0$ plane). Here the current density is normalized by local magnetic field strength, which provides a high contrast of thin current layers with other volumetric currents. {The unit of $J/B$ is $1/\Delta$ where $\Delta$ is the grid size.}
  (b) SDO/AIA images of the X8.2 flare observed at the solar limb. The images are made by combination of two AIA channels 211~{\AA} and 304~{\AA}, and they are rotated to roughly match the direction of the simulated eruption.
  (c)-(d) Distributions of magnetic squashing degree $Q$ and twist number $T_w$, respectively, on the same cross section in (a).}
  \label{fig4}
\end{figure*}

%2D structure as shown in a cross section
The {eruption} is characterized by a drastic rise and expansion of the MFR in the MHD simulation, which is shown in \Fig~\ref{fig4} for a vertical cross section. As clearly seen from the cross section of current density ($J/B$, \Fig~\ref{fig4}a), a narrow current layer of upside-down teardrop shape forms at the boundary the MFR. Such a boundary is precisely depicted by a QSL, as shown in the map of magnetic squashing degree (\Fig~\ref{fig4}c), and the boundary becomes more and more distinct as the MFR expands with time. Such expansion is also reflected in the evolution of region with strong twist number (\Fig~\ref{fig4}d). The rising path of the MFR deviates from the vertical towards the east (the $-x$ direction), where the magnetic pressure is weaker than the west (the $+x$ direction).

\begin{figure*}[htbp]
  \centering
  \includegraphics[width=0.8\textwidth]{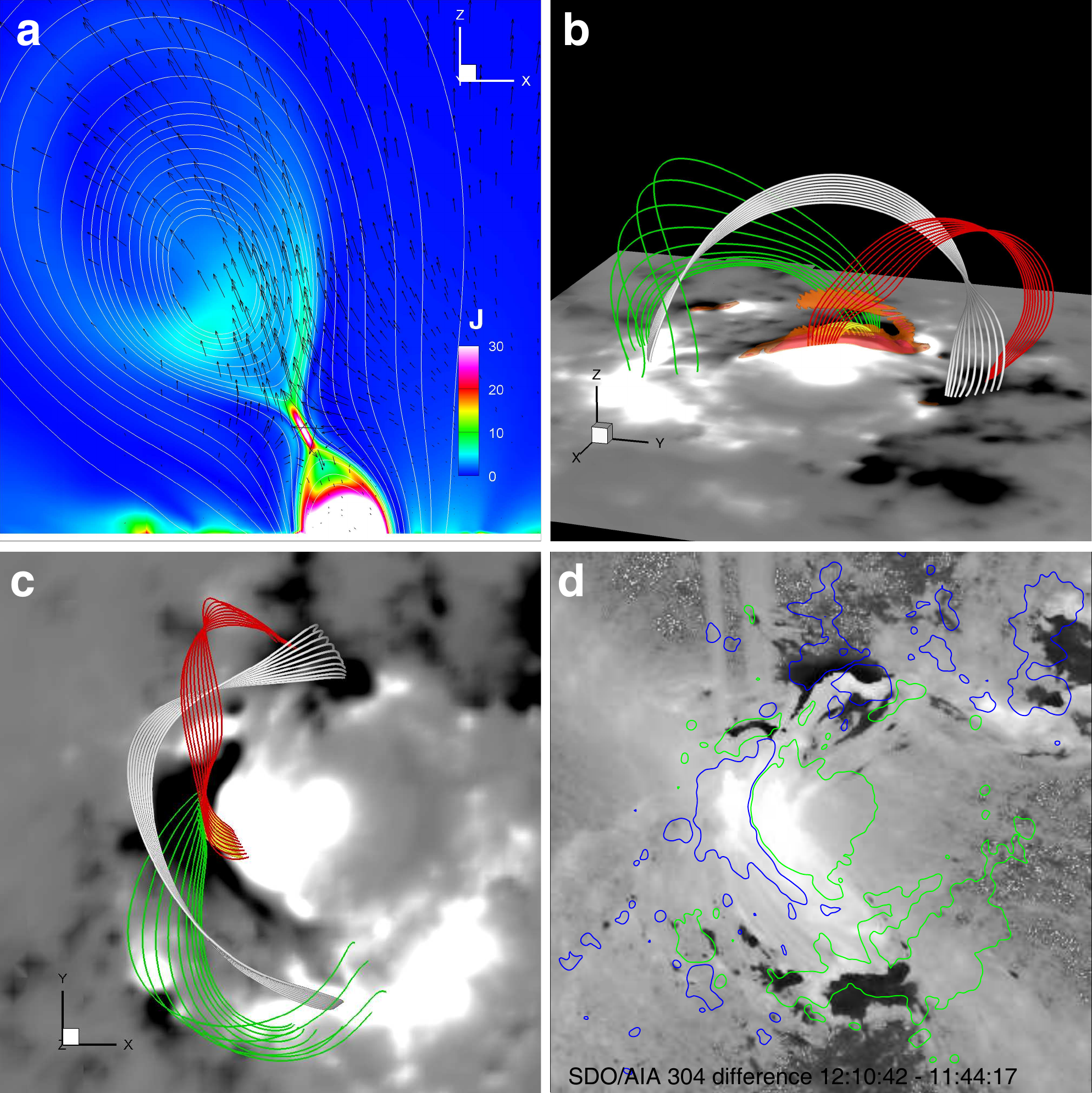}
  \caption{
  {Illustration of the reconnection process below the rising
    MFR.} (a) Current density distribution
    and plasma flows (denoted by arrows) on the vertical cross section (the $y=0$ plane) at time of $t=1.0$. The white lines are 2D field lines tracing
    on the plane. (b) 3D configuration of the reconnection.
    The white lines {are within} the main body of the
    MFR. The red and green lines {are} reconnecting
    field lines below the rope.
    Their inner footpoints are sheared past each along the PIL,
    and thus the field directions change abruptly across the CS, in which reconnection takes
    place, results in a long field line joining the MFR and a short
    arcade below which forms the post-flare loops (as shown by the yellow lines). The objects colored in
    red and orange are thin layers with strongest current density throughout the volume, showing the CS in 3D.
    (c) Top view of the same magnetic field lines shown in (b).
    (d) Difference of AIA 304~{\AA} images of post-flare with pre-flare time showing two dimming sites (or transient coronal holes) that match the locations of the MFR legs. The contour lines are shown for $B_z = -500$~G (colored in blue) and $500$~G (colored in green).}
  \label{fig6}
\end{figure*}

Starting from the initial BP point, there evolves an intersection of QSL below the MFR, forming an X shape of increasing height and size. Such QSL intersection is a magnetic null-point configuration in the 2D plane, while in 3D it is known as an HFT, where highest $Q$ values are often found, making the HFT a preferential site for current accumulation and subsequent dissipation through magnetic reconnection~\citep{Titov2002}. Indeed, an intense CS forms (with {$J/B > 0.2/\Delta$, where $\Delta$ is the grid size}) there (compare \Fig~\ref{fig4}a and c), and it also evolves into an X shape. Moreover, magnetic reconnection is triggered in the HFT (or the CS) as indicated by the plasma flows near the CS (see \Fig~\ref{fig6}a), which shows a typical pattern of reconnection flows in 2D, i.e., bi-directional horizontal inflows at two sides of the CS and bi-directional vertical outflows away from the X point. This reconnection along with the cusp-CS-rope configuration (see the 2D field lines in \Fig~\ref{fig6}a) reproduce nicely the picture of the standard flare model in 2D. Interestingly, the shape of enhanced current layer and its evolution look rather similar to the AIA images of the limb X8.2 flare in the same AR~(\Fig~\ref{fig4}b), although the simulation is not aimed for that flare. The X8.2 flare is characterized by a bright ring enclosing a relatively dark cavity of increasing size. The coronal cavity often indicates an MFR \citep{GibsonFan2006}, while its outer edge is bright because heating is enhanced there by the dissipation of the strong current in the boundary layer of the rope. Below the cavity is an even brighter cusp-shaped flare loop system, the shape of which is also seen in the current distribution in the simulation.

\begin{figure*}[htbp]
  \centering
  \includegraphics[width=0.6\textwidth]{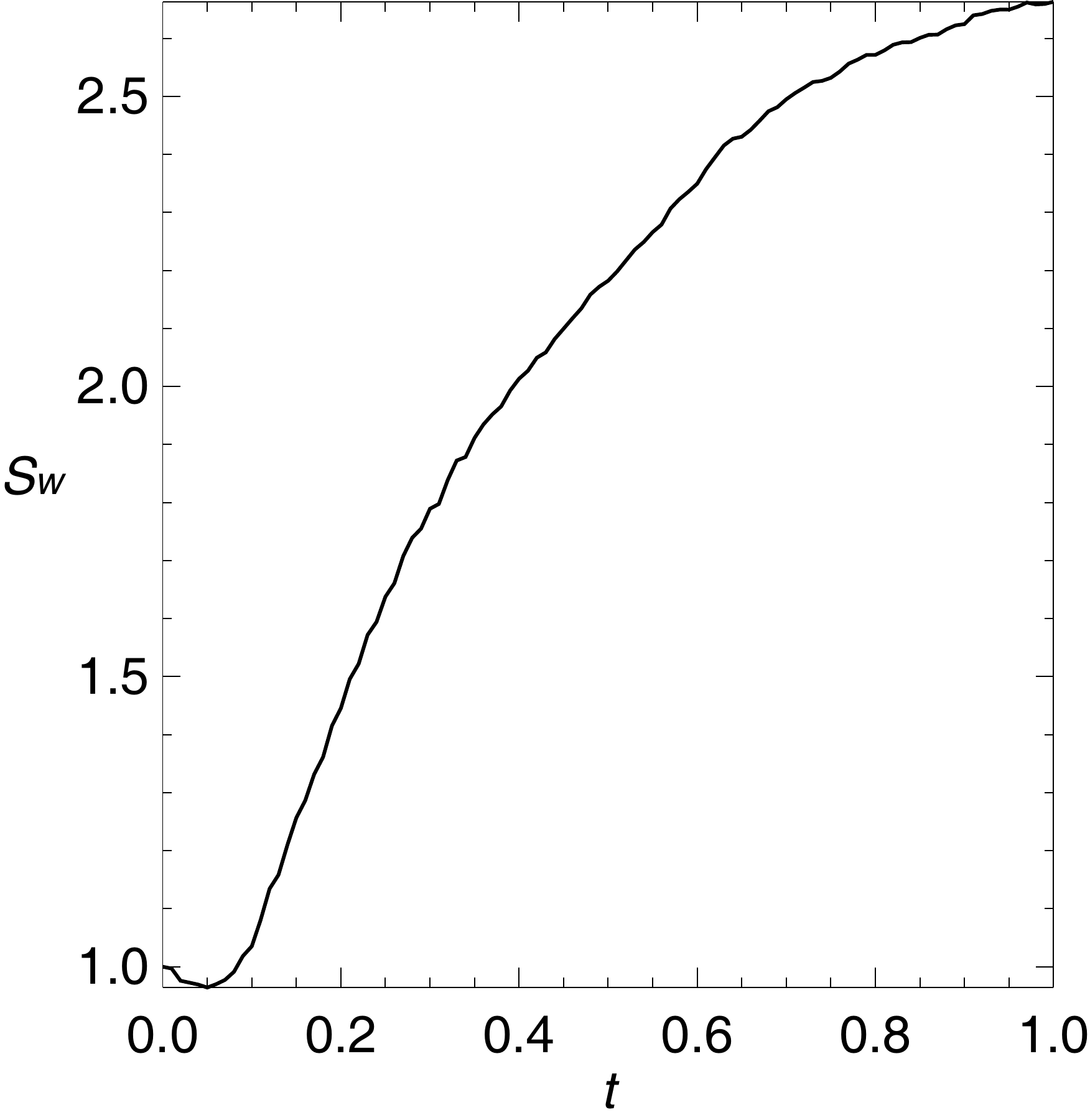}
  \caption{
  {Evolution of twisted flux content.} The magnetic flux content $S_w$ of the MFR is calculated as $S_w = \int |T_w B_y| dx dz$ on the surface $y=0$, and the integration is limited to the area of $|T_w|>1$. The number of $S_w$ as shown is normalized by its initial value at $t=0$.}
  \label{fig4.1}
\end{figure*}

Evolution of the magnetic twist distribution as shown in \Fig~\ref{fig4}d indicates that the reconnection adds magnetic twist to the MFR~\citep{WangW2017}. In \Fig~\ref{fig4.1}, we show the time evolution of the sum of the magnetic flux content (multiplied by twist number) of the MFR with twist number above unity. Significant increase of twisted flux with time can be seen. The twist is added to the outer layer of the rope (\Fig~\ref{fig4}d), while the central part of the rope maintains the pre-flare twist numbers.

\begin{figure*}[htbp]
  \centering
  \includegraphics[width=0.8\textwidth]{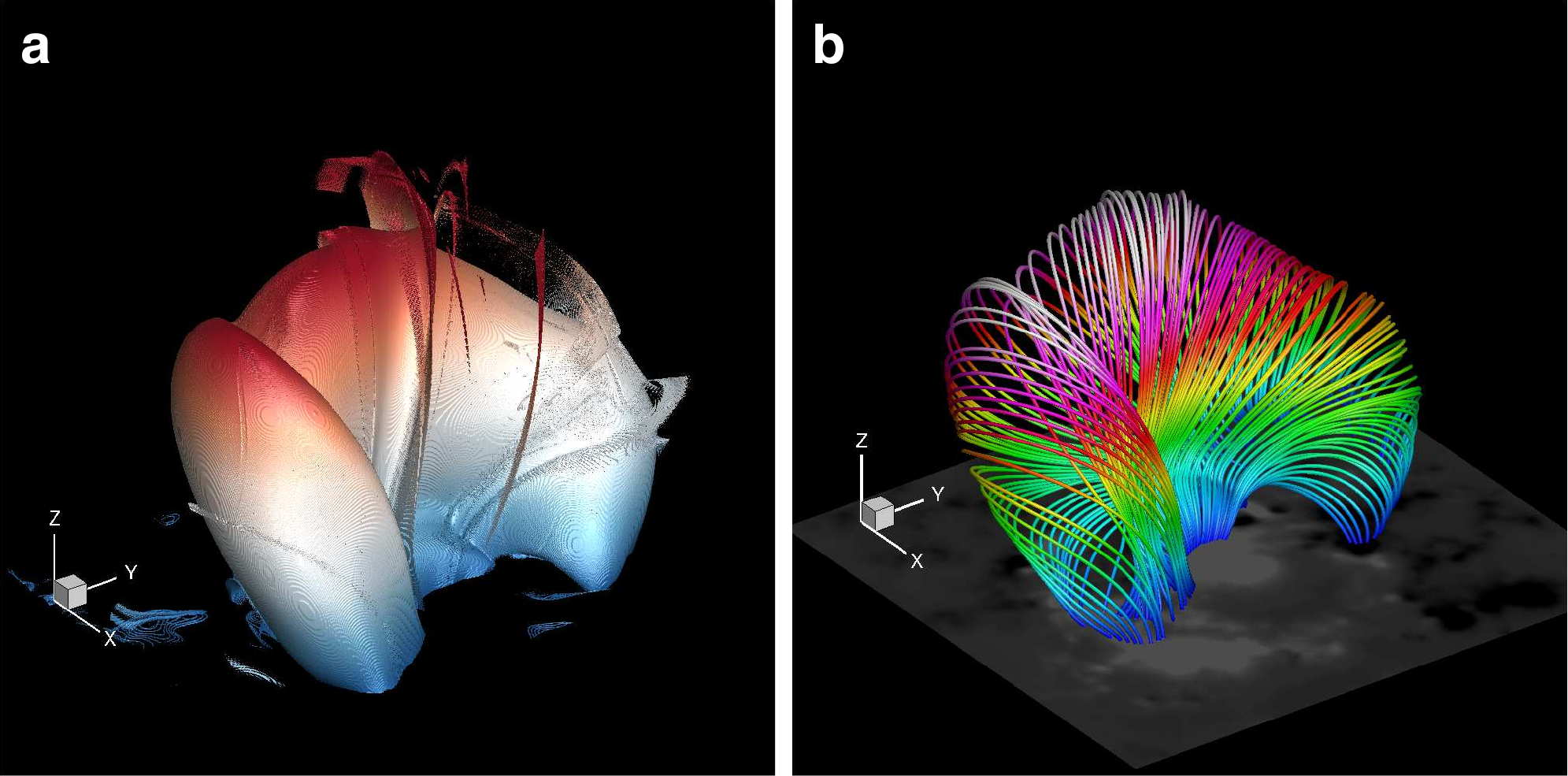}
  \caption{
  {3D structure of the QSL that wrap the erupting MFR at simulation time $t=1$.} (a) The iso-surface of $Q=1000$, (b) Sampled magnetic field lines that form the QSL. The colors denote the value of height $z$. A animation of rotating view of the structure is provided in Supplementary movie~1, and a moving slice crossing through the MFR in $y$ direction, which shows the structure of the the QSL and the twist number in the MFR, is provided in Supplementary movie~2.}
  \label{fig6.1}
\end{figure*}

\begin{figure*}[htbp]
  \centering
  \includegraphics[width=\textwidth]{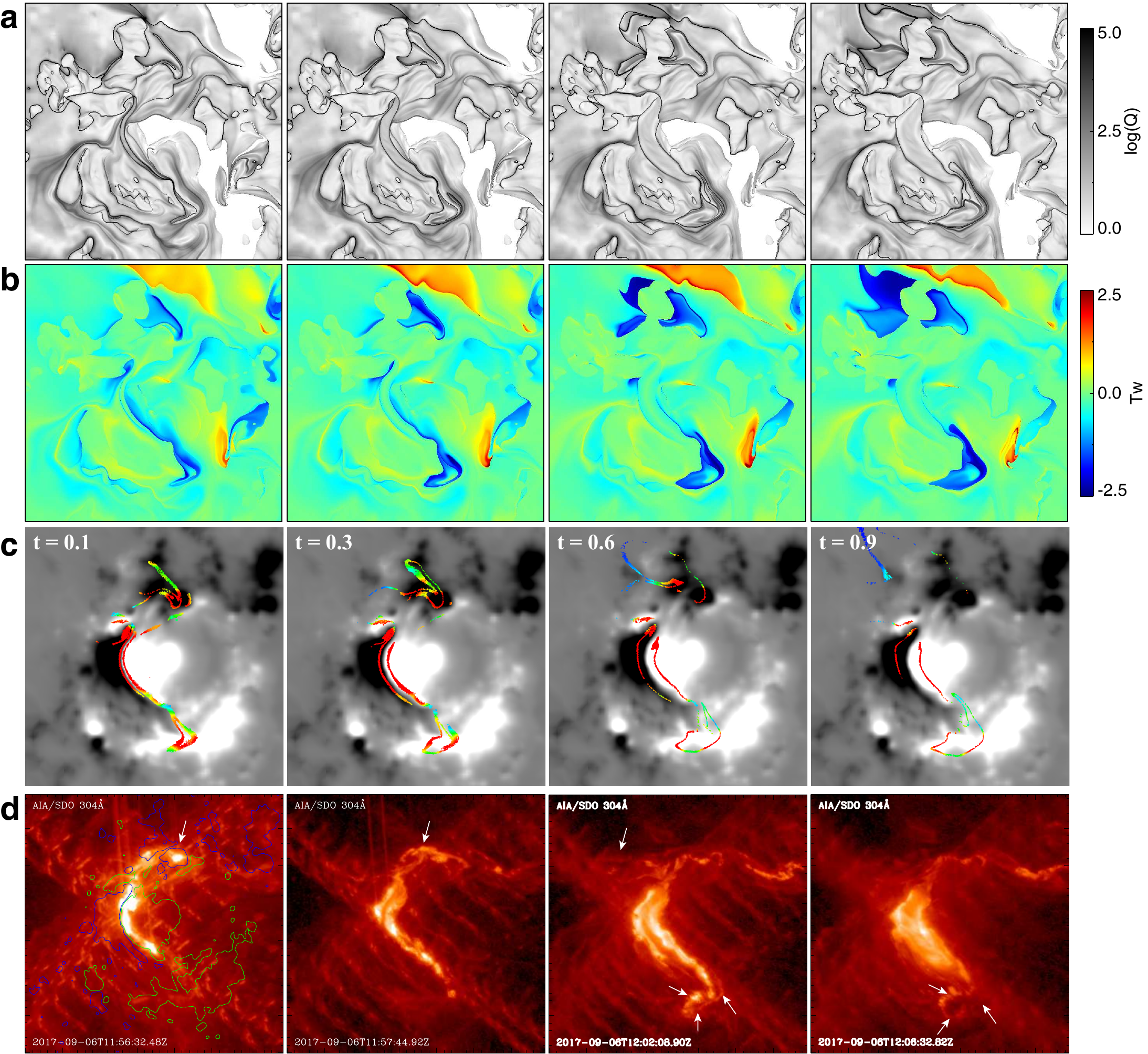}
  \caption{
  {Structures and evolution at the bottom surface.}
  (a) Magnetic squashing degrees. (b) Magnetic twist numbers. An animation of the temporal evolution of the magnetic squashing degree and twist number is provided in Supplementary movie~3. (c) Magnetic footpoints (color dots) of the field lines that are traced from the CS to the bottom surface. The colors represent the strength of local magnetic field, red for strong and blue for weak. (d) SDO/AIA~304~{\AA} images of the flare ribbons.  The arrows denote the two weak ribbons that form in the far-side polarities P1 and N1. An animation of the flare ribbon evolution is provided in Supplementary movie~4. The ribbons as observed in this channel look rather {diffuse}, but the AIA UV channels of 1600~{\AA} and 1700~{\AA} are overexposed in this flare.}
  \label{fig7}
\end{figure*}

%3D illustration.
In 3D, the MFR's surface (or boundary) is rather complex, but an arched tube structure can be seen from the 3D QSLs~(\Fig~\ref{fig6.1} and Supplementary Movie~1), within which the magnetic twist is distinctly stronger than that of the ambient flux~(Supplementary Movie 2, see also the QSL and twist distribution on the bottom surface in \Fig~\ref{fig7}a and b). With the rising of the MFR body, its
conjugated legs are rooted in the far-side polarities P1/N1, while its pre-flare connections to P0 and N0 is cut by the reconnection. Consequently, the initial elongated distribution of magnetic twist along the main PIL becomes coherent in the two feet of the rope, which expand in size with time (\Fig~\ref{fig7}b and Supplementary Movie~3). Both feet are rather irregular: while the southern one has a high-$Q$ boundary, which is highlighted by brightening in AIA~304~{\AA} (\Fig~\ref{fig7}d), the northern one is even more complex, which initially has a closed boundary but quickly splits into two fractions due to the mixed magnetic polarities there. As can be seen in Supplementary Movie~3, the south foot expands from the initial closed QSL that separate the P1-N1 flux from those of other connections. Thus the initial P1-N1 flux provides a seed for the subsequent erupting rope, and the weakly-twisted seed flux remains in the rope's core, which is surrounded by highly-twisted flux.

\begin{figure*}[htbp]
  \centering
  \includegraphics[width=\textwidth]{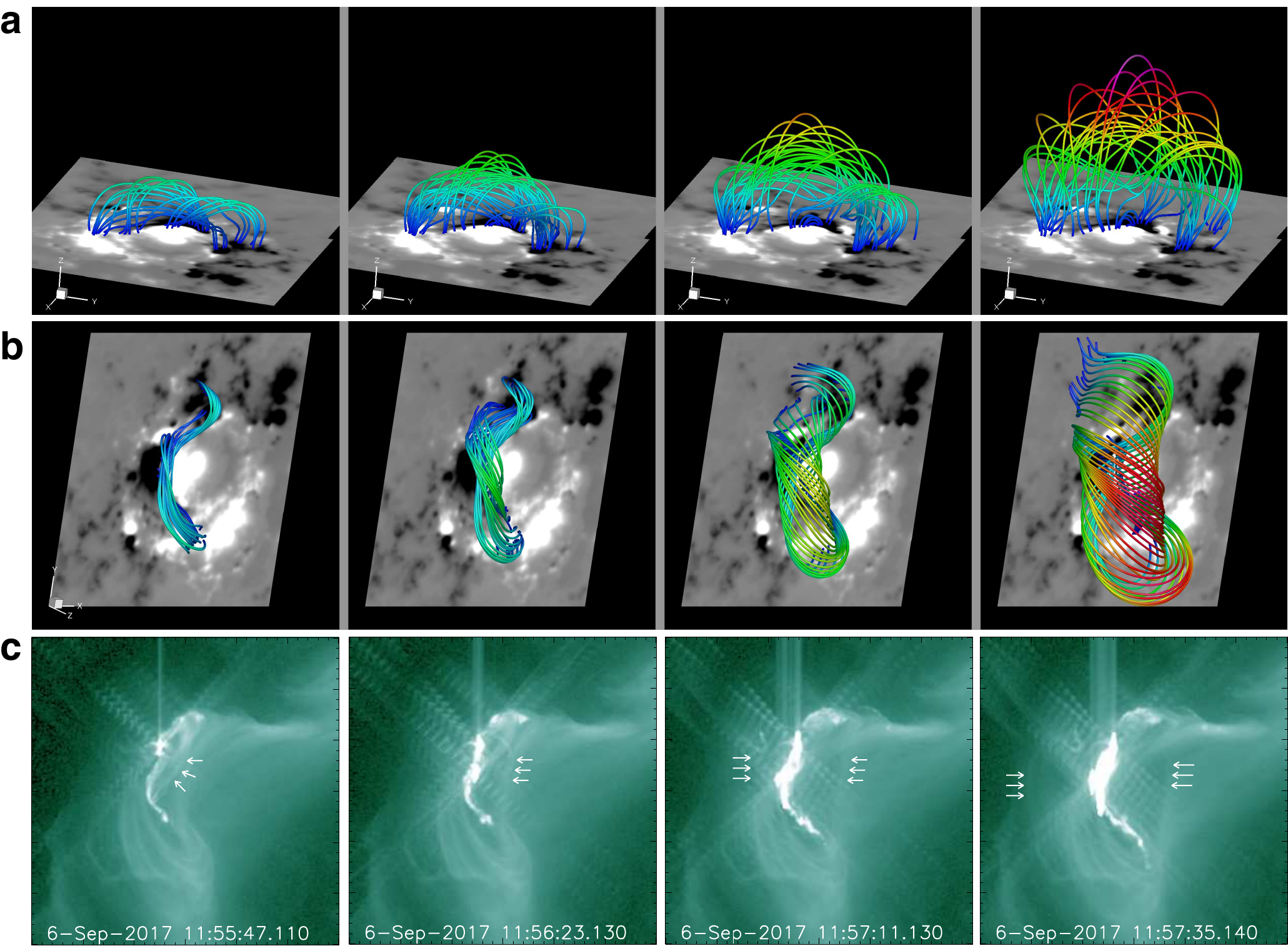}
  \caption{
  {The eruptive structure in 3D and comparison with SDO/AIA observation.}
  (a) Side view of sampled magnetic field lines of the erupting MFR. The magnetic field lines are false-colored by the value of height $z$ for a better visualization. The bottom surface is shown with the photospheric magnetogram.
  (b) Magnetic field lines that form the surface of the MFR (see Supplementary movie~1 for a 3D rotation view of these field lines, as well as the 3D structure of the QSL.). The view angle is arranged to be the same as that of the SDO.
  (c) SDO/AIA~94~{\AA} observations of the erupting process. Two sets of arrows mark the two expanding edges, presumably corresponding to the expanding surface of the MFR. Such expanding features can be seen more clearly in the animation provided in Supplementary Movie~5.}
  \label{fig5}
\end{figure*}

The 3D magnetic field lines of the evolving MFR is shown in \Fig~\ref{fig5}a. The expansion of the tube-like flux rope can be visualized by the evolution of the magnetic field lines traced from the rope's high-$Q$ boundary (\Fig~\ref{fig5}b). This expansion find its signatures in an EUV hot channel observed by SDO/AIA~94~{\AA} from above (see \Fig~\ref{fig5}c). Two bright edges are observed to expand away from the main PIL (see also the Supplementary Movie~5), which agrees well with the expanding surface of the simulated rope (as seen in the same view angle, \Fig~\ref{fig5}b). Such EUV hot channels are often deemed to be associated with erupting MFRs~\citep{ZhangJ2012, Cheng2012}, and here it is suggested to correspond specifically to the surface or topological interface of the MFR. There appears no writhing signature of the rope, as seen in both the simulation and the observation, indicating that the KI did not occur. Thus the TI is probably the driving mechanism of the eruption, at least in the early stage of this eruption.

After the eruption onsets, the footprint of the BPSS, which is aligned along the main PIL, starts to bifurcate (\Fig~\ref{fig7}a and Supplementary Movie~3). The bifurcation nicely matches the two parallel flare ribbons departing the PIL (\Fig~\ref{fig7}d and Supplementary Movie~5). This is associated with the transformation of the BPSS into the HFT, as the MFR rises and develops into a coherent structure (Figure~\ref{fig4} and \ref{fig5}). The dynamic flare ribbons reflect the instantaneous footpoints of magnetic field lines undergoing reconnection. These field lines form the QSL boundary of the MFR and pass through the CS (or the HFT) below the rising rope. Thus, the flare ribbons, the footpoints of field lines threading the CS, and the footprint of the HFT are co-spatial, as demonstrated in Figure~\ref{fig7}. The reconnection in 3D occurs in a tether-cutting-like fashion. As illustrated in \Fig~\ref{fig6}b and c, the reconnecting field lines (red and green) are highly sheared and their inner footpoints below the rope (white) are close to each other, where the field directions change abruptly across the CS. The reconnection will produce a long twisted field line winding the rope and a short post-flare loop (yellow), therefore `cutting loose' the sheared field lines from their original anchors below the rope.

\begin{figure*}[htbp]
  \centering
  \includegraphics[width=0.8\textwidth]{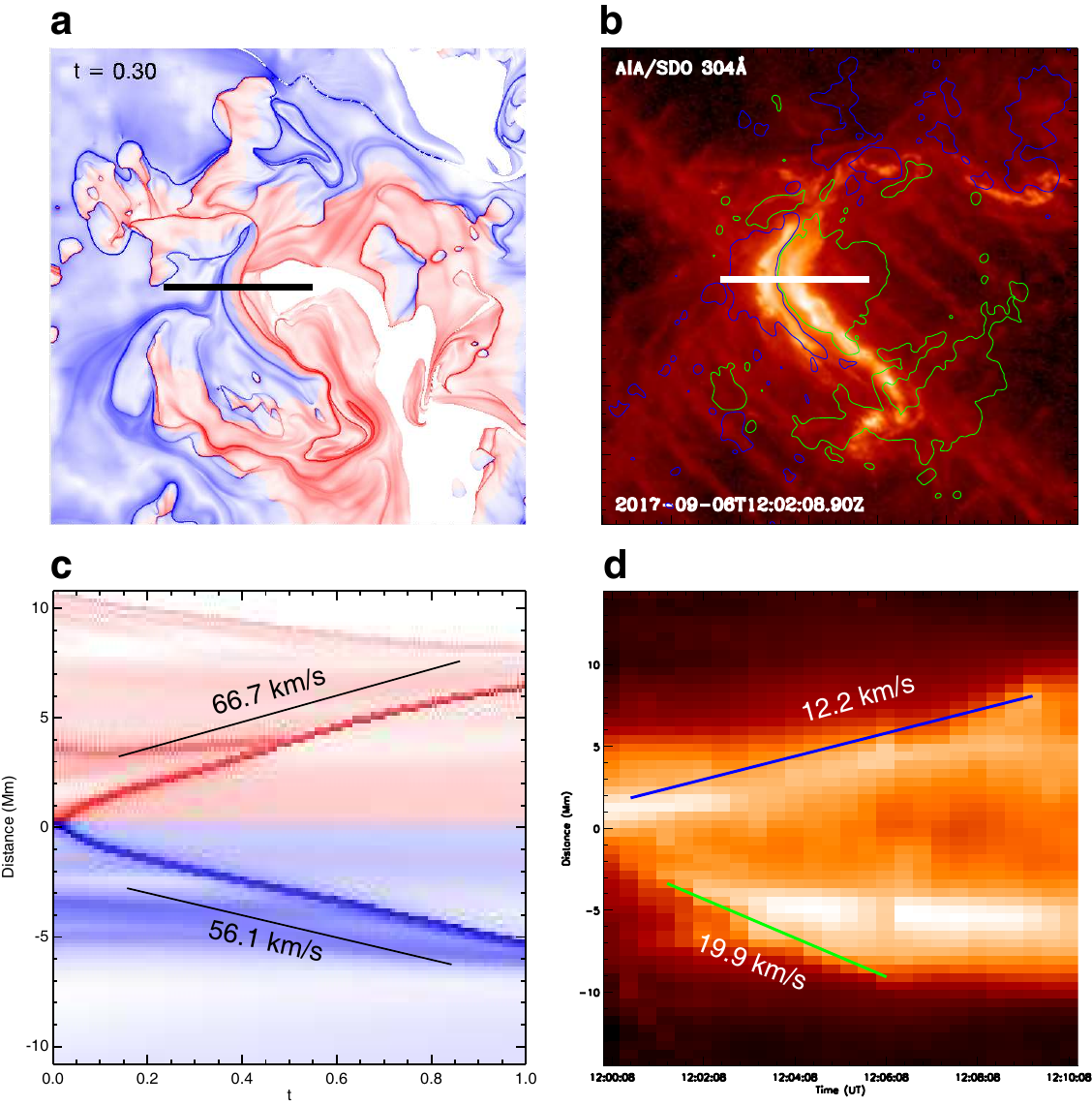}
  \caption{
  {Comparison of flare-ribbon separation speed from MHD simulation and that from observation.}
  (a) The QSL map on the bottom at time of $t=0.3$. The black line segment denotes the location where a time sequence of stack is plotted in (c) for tracking the separation of the main QSLs that maps the reconnection footpoint. (b) SDO/AIA 304~{\AA} image of the flare ribbons. In the same way, the white line segment is the location of the time stack shown in (d), which shows the separation motion of the main flare ribbons. The lines shown in (c) and (d) represent approximately the speeds of the ribbons with time, respectively.}
  \label{Sfig1}
\end{figure*}

In \Fig~\ref{Sfig1}, the separation speed of the two main ribbons from the simulation is compared with that from observation: the modeled speed, $50 \sim 70$~km~s$^{-1}$, is approximately $3\sim 5$ times faster than the observed one ($10\sim 20$~km~s$^{-1}$).
The reconnection rate, as measured by the ratio of inflow speed to the local Alfv{\'e}n speed (i.e., the inflow Alfv{\'e}n Mach number), is $0.05  \sim 0.1$ in the MHD simulation, which is comparable to the estimated values from direct observation analysis of various flares~\citep{SuY2013, SunJ2015}. Thus, the real value of the reconnection rate for this X9.3 flare should be $0.01\sim 0.03$, if we divide the simulated reconnection rate by the ratio of the simulated ribbon separation speed to the observed one.
\textbf{It is plausible that our model produces reconnection much faster than the observed rate because the numerical resistivity is much larger than that of the real corona (see Section~\ref{sec:mhd})}.

\begin{figure*}[htbp]
  \centering
  \includegraphics[width=0.8\textwidth]{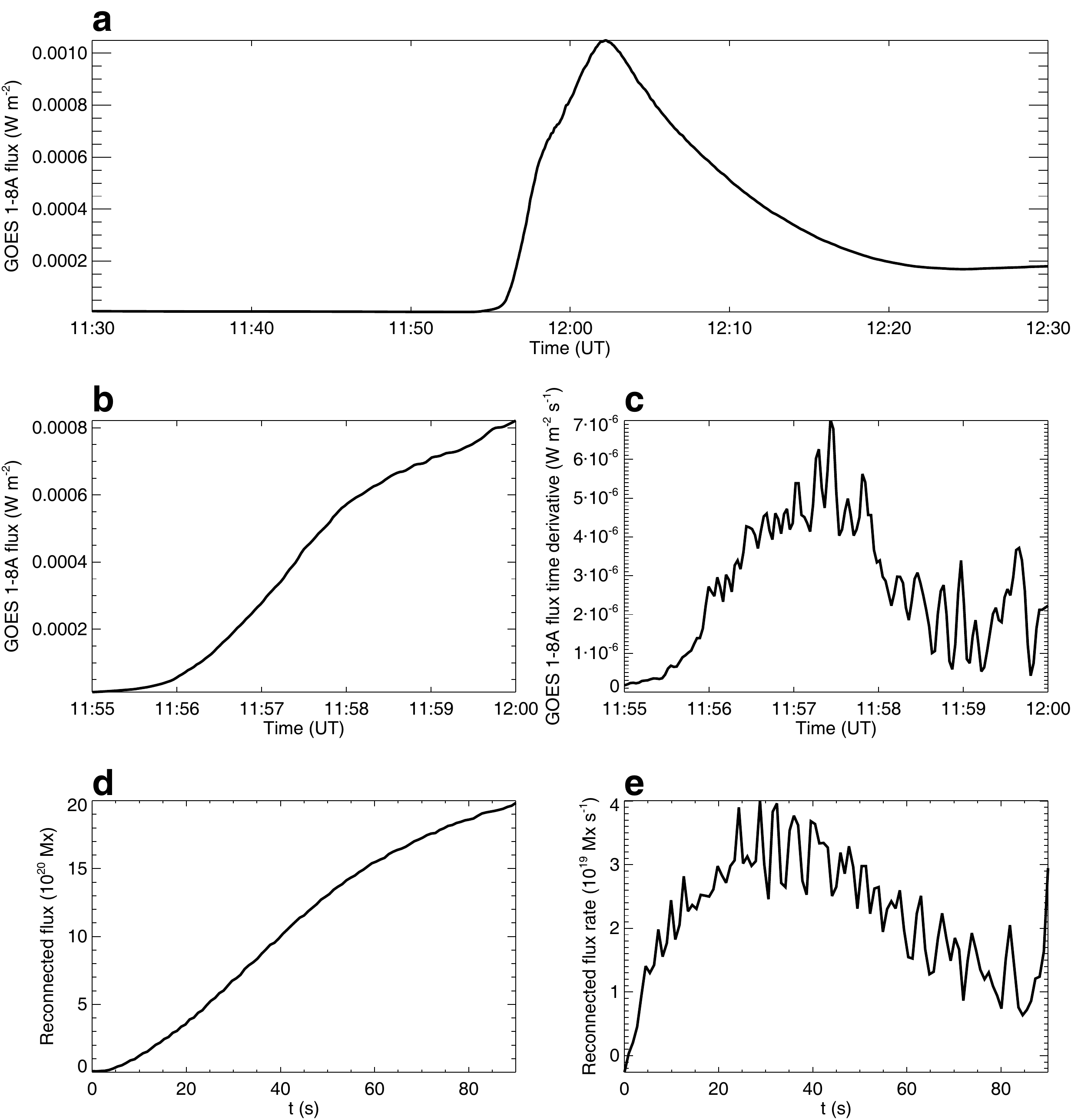}
  \caption{
  {Reconnected flux as compared with profile of X-ray flux. }
  (a) The GOES soft X-ray flux (1-8~{\AA}) from 11:30 to 12:30, September 6, 2017. (b) Same as (a) but for a small interval of 5~min starting from the flare beginning. (c) The time derivative of the flux shown in (b). (c) The temporal evolution of reconnected flux in the simulation. (d) Time derivative of the reconnection-flux curve, i.e., reconnection flux rate.}
  \label{reconflux}
\end{figure*}

In the simulation, the reconnection flux can be precisely calculated, which is the magnetic flux swept by the main ribbons of the QSL footprints. In \Fig~\ref{reconflux}, we compare the temporal evolution of the reconnection flux with GOES soft X-ray flux in 1-8~{\AA}, which can be used to represent the time profile of the flare energy release. In the simulation of one time unit $\tau=90$~s, a total flux of $2\times 10^{21}$~Mx reconnects, and the temporal rate of the reconnection flux impulsively reaches its peak at $\sim 30$~s, and then decreases. It can be seen that the profile of soft X-ray flux from 11:55~UT to 12:00~UT is comparable with that of the reconnection flux~(compare \Fig~\ref{reconflux}b and d). The ratio of the observed time (5~min) to our simulated one (90~s), is also consistent with the ratio of the simulated ribbon separation speed to the observed one. Furthermore, the reconnection rate profile looks rather similar to the time derivative of the X-ray flux (compare \Fig~\ref{reconflux}c and e), despite that the latter increases less rapidly than the simulated one. This indicates that the simulation reproduced the early process of flare-energy impulsive release, except that the numerical model enhanced the reconnection rate by several times.    %Besides, our MHD model cannot reproduce a precise mechanism of the reconnection process, which is still unclear.
%It should be noted that here the comparison in temporal evolution with the observation can only be made qualitatively, because here the MHD model does not include the precise mechanism of reconnection and the rate of reconnection cannot be reproduced.

In addition to the double ribbons parallel to the main PIL, relatively weak ribbons extend to the two remote polarities, P1 and N1, forming a nearly closed ring at P1, similar to the observation in \citet{WangW2017}, while its counterpart at N1 is rather complex. These ribbons are produced by reconnection-released energy depositing at the far ends of the reconnecting field lines threading the CS, therefore constituting the boundary of the MFR's feet, which expands with the separation of the double ribbons (Figure~\ref{fig7}). The expansion is attributed to reconnections at the CS, which add successive layers of twisted flux to the MFR. In observation, the feet of an erupting MFR are often indicated by a pair of transient coronal dimming, resulting from plasma evacuation along the MFR legs into interplanetary space~\citep{Qiu2007, Webb2000}. In \Fig~\ref{fig6}d, an AIA 304~{\AA} image taken after the eruption is subtracted by one before the eruption, showing two distinct patches of coronal dimming at N1 and P1, which match the two feet of the modeled MFR.

%[this part will be left for a future study] Finally, the modeled post-flare loops, which is the short magnetic arcades immediately below the flare CS, show a strong-to-weak transition of magnetic shear with the progress of the flare reconnection (\Fig~\ref{fig8}a). Such configuration evolution of flare loops is well observed for typical two-ribbon flares~\citep[e.g.,][]{SuY2006, SuY2007, Aulanier2012, Vemareddy2018}. Another well-documented fact we have reproduced (\Fig~\ref{fig8}b) is a permanent enhancement of the transverse field along the main PIL on the bottom surface after flare~\citep[e.g.,][]{WangH2012}.

\section{Conclusion}
\label{sec:concl}

In this paper, through a combination of observation data and numerical simulation, we have revealed the topological evolution of magnetic configuration associated with a great eruptive flare, the largest one in solar cycle~24. Reconstruction of the coronal magnetic field immediately prior to the flare results in an complex MFR system that consists of multiple bundles of field lines with different connections and twist degrees. Owing to a strongly distorted, quadrupolar photospheric magnetic configuration, the main body of the MFR forms a C shape, unlike {more typically} observed sigmoidal one. Magnetic field lines of the MFR run horizontally over the strongly sheared PIL in the core of the AR. The bottom of the rope is attached on the photosphere, resulting in BPs along the PIL. Analysis of the decay index of the background potential field in the vicinity of the MFR shows that a major part of the MFR already enters into the TI domain.

The unstable nature of the pre-flare magnetic field results in fast expansion and rising of the MFR in the MHD simulation as initialized by the reconstruction data. In the wake of the rising MFR, an HFT comes into being from the BPs, resulting in an intersection of the QSL that warps the rope. Strong current density thus accumulates in the HFT, forming a CS and reconnection is consequently triggered there. Magnetic twist is sequentially built up on the outer layer of the rope through reconnection of the field lines there. The modeled magnetic configuration and evolution are found to be consistent with observed EUV features of the eruption, such as the expanding hot channels that are presumably resulted by the enhanced emission in the MFR-related QSL, the dark cavity with bright edge that corresponds to the cross section of the rope (although imaged for another flare of the same AR), and the coronal dimming in the feet of the rope. Most importantly, by tracing the newly-reconnected field lines from the CS to the bottom surface, we have reproduced the location of two main flare ribbons as well as their separation. We estimated the average reconnection rate of the flare to be $0.01\sim 0.03$ by comparing the simulated ribbon separation speed with the observed one. Furthermore, the temporal profile of the simulated reconnection flux is comparable to the observed soft X-ray flux. In addition to the main ribbons, there are relatively weak flare ribbons of close shape extending to the two feet of the rising rope, which is successfully matched by the far-end footpoints of the newly-reconnected field lines or the footpoints of the MFR-related QSL. The areas enclosed by these ribbons increase gradually as increasing amount of flux joins the MFR through the reconnection.

The significance and also uniqueness of our simulation is that we did not use any prior assumption on the magnetic configuration of the eruptive structure. The pre-flare flux-rope complex is reconstructed from directly the vector magnetogram and its evolution to a coherent MFR is self-consistently reproduced by MHD model. This is distinct from many other data-constrained simulations of solar eruption, which are made based on the prior assumption of the existence of MFR. For example, in a series of papers~\citep{Savcheva2015, Savcheva2016, Janvier2016}, an MFR with its axis roughly fitting the observed filament is inserted into a potential field environment, and consequently the bottom boundary does not match the observed magnetogram. In the works of~\citet{Inoue2015, Inoue2018}, an MFR is made by reconnection between sheared magnetic arcades that is reconstructed by NLFFF model of the pre-flare corona. {By including an ad-hoc, current-dependent anomalous resistivity, reconnection occurs above the strong-sheared neutral line. It is unclear whether the flux rope and subsequent flare reconnection can be reproduced without the anomalous resistivity in their model.}
{However, different from simulations of~\citep{Savcheva2016, Inoue2018, Torok2018} which explicitly alter a stable initial field to make it unstable, here our simulation started from an already unstable field. Thus it should be noted that, because we have not investigated whether any pre-flare NLFFF configuration would be stable in our dynamic MHD model, we cannot rule out the possibility that the vector magnetograms significantly prior to (for instance, hours before) the actual flare might also produce an eruption in our model. In other words, the occurrence of an eruption in our model might not depend upon details of the input magnetogram. This issue needs to be investigated in future works for the purpose of identifying the true mechanisms triggering eruptions.}

In summary, without any prior assumption on the magnetic configuration, we reproduced the eruptive process of a great solar flare with numerical MHD simulation based entirely on a pre-flare vector magnetogram on the photosphere. Without the simulation, it would be extremely difficult to envision the evolving magnetic topology, although the evolution is hinted by observational features, e.g., the originally incongruous MFR growing into a coherent structure via magnetic reconnection, which is manifested in the evolving ribbon morphology that significantly deviates from two parallel or J-shaped ribbons in the standard picture. To conclude, such a realistic model and comprehensive analysis provide a sophisticated characterization of the invisible coronal magnetic field behind the eruption, which is of utmost importance for the eventual understanding and forecast of solar flares/CMEs.

%===========================================================
\acknowledgments
Data from observations are courtesy of NASA {SDO}/AIA and the HMI
science teams. The computation work was carried out on TianHe-1 (A) at
the National Supercomputer Center in Tianjin, China.
This work is supported by the National Natural Science Foundation of
China (NSFC 41822404, 41731067, 41574170, 41531073), the Fundamental Research Funds for the Central Universities (Grant No. HIT.BRETIV.201901), and Shenzhen Technology Project JCYJ20170307150645407. C.J. and Q.H. also acknowledge NSF grant AGS-1650854 and NASA grant 80NSSC17K0016 for partial support.
R.L. acknowledge support by NSFC 41474151, 41774150, and 41761134088.
P.V. acknowledges support by an INSPIRE grant of AORC
scheme under the Department of Science and Technology of India. We are very grateful to the anonymous referee for his meaningful
comments and suggestions.

%\bibliographystyle{apj}
%\bibliography{all}

%--Fig1 show location of flare

\end{CJK*}
\end{document}